\documentclass[prl, twocolumn,amsmath,amssymb, superscriptaddress,longbibliography, nofootinbib]{revtex4-1}
\usepackage[pdftex,colorlinks=true]{hyperref}
\usepackage{graphicx}% Include figure files
\usepackage{natbib}
\usepackage{amsmath,epsfig,color,amssymb}
\usepackage{bbold}
\usepackage[dvipsnames]{xcolor}
\usepackage[export]{adjustbox}
\usepackage{comment}
\newcommand{\beg}{\begin{equation}}
\newcommand{\en}{\end{equation}}

\newcommand \bel  {\begin{align}}
\newcommand \enl  {\end{align}}

\definecolor{new}{rgb}{.08,.05,.8}

\begin{document}
\author{Areg Ghazaryan}
\affiliation{Institute of Science and Technology Austria, Am Campus 1, 3400 Klosterneuberg, Austria}

\author{Ammar Kirmani}
\affiliation{Physics Department, City College of the City University of New York, NY 10031, U.S.A.}

\author{Rafael M. Fernandes}
\affiliation{School of Physics and Astronomy, University of Minnesota, Minneapolis, Minnesota 55455, US}

\author{Pouyan Ghaemi}
\affiliation{Physics Department, City College of the City University of New York, NY 10031, U.S.A.}
\affiliation{Physics Program, Graduate Center of City University of New York, NY 10031, U.S.A.}

\title{Anomalous Shiba states in topological iron-based superconductors}

\begin{abstract}
We demonstrate the formation of robust zero energy modes close to magnetic impurities in the iron-based superconductor FeSe$_{1-x}$Te$_x$. We find that the Zeeman field generated by the impurity favors a spin-triplet inter-orbital pairing as opposed to the spin-singlet intra-orbital pairing prevalent in the bulk. The preferred spin-triplet pairing preserves time-reversal symmetry and is topological, as robust, topologically-protected zero modes emerge at the boundary between regions with different pairing states. Moreover, the zero modes form Kramers doublets that are insensitive to the direction of the spin polarization or to the separation between impurities. We argue that our theoretical results are consistent with recent experimental measurements on FeSe$_{1-x}$Te$_x$.

%The experimental puzzles which we try to address:
%\begin{enumerate}
%    \item The presence of zero energy mode close to Fe impurity \cite{yin2015observation,fan2021observation}
%    \item The zero mode does not split under up 8T magnetic field \cite{yin2015observation,fan2021observation}
%    \item The zero mode does not split when two impurities are close to each other \cite{yin2015observation}
%    \item The zero energy mode splits when close to the vortex with zero energy \cite{fan2021observation}
%    \item Conversion of the usual Shiba state into zero mode when bringing tip closer to the atom (possibly increasing the exchange interaction) \cite{fan2021observation}
%    \item Zero mode is not spin-polarized \cite{wang2021spin}
%    \item Zero mode disappears when the impurity is moved  away from $C_4$ point \cite{fan2021observation}
%    \item Energies of in gap states are integer quantized \cite{fan2021observation} 
%\end{enumerate}
\end{abstract}

\maketitle

\textit{Introduction:} Topological superconductivity (TSc) is a quantum state {that has been extensively explored} ~\cite{hasan2010colloquium,qi2011topological,sato2017topological}, particularly due to its application for realizing Majorana zero modes (MZM)~\cite{Alicea_2012,Flensberg2021}. In recent years, the iron-based superconductor FeSe$_{1-x}$Te$_x$ (FST) emerged as a promising candidate for TSc ~\cite{wang2015topological}. 
It was theoretically predicted that the band inversion in the electronic structure of FST can lead to localized MZM at the end of vortex lines in the superconducting phase~\cite{xu2016topological}. One of the appeals of superconducting FST is the comparable energy scales between the superconducting gap and the Fermi energy, which leads to a wide spectral resolution of the vortex zero mode~\cite{kreisel2020remarkable}. The experimental observation of such vortex zero modes has led to extensive studies of FST as a bulk TSc~\cite{zhang2018observation,wang2018evidence,zhu2020nearly,machida2019zero,miao2018universal}. {The interplay between topology, superconductivity and magnetism has also been investigated in relation to surface states and generating new superconducting order \cite{li2021electronic,hu2020pairing,zaki2021time,youmans2018odd,dzero2022impurity,li2022topological}}. On the other hand, several of the properties of FST, such as the existence of vortices with and without zero modes in the same sample, have been puzzling~\cite{machida2019zero,Vidya,chiu2020scalable,ghazaryan2020effect,ruixing}. 

Another type of in-gap states in superconductors are the Shiba states, which form near magnetic impurities~\cite{shiba1,shiba2,shiba3}. {In certain regimes in-gap states can also be formed for non-magnetic impurity \cite{sau2013bound} even when the system respects time-reversal symmetry \cite{matsumoto2009single,mashkoori2019impact}} In FST, in-gap states have been observed near interstitial Fe atoms~\cite{yin2015observation}, which behave as magnetic impurities~\cite{PhysRevLett.108.107002}.  Interestingly, several of the properties of these in-gap states are different from those of conventional Shiba states. First, Shiba states generally have a finite energy unless the microscopic properties of the superconductor and the coupling strength of the magnetic impurity with the itinerant electrons are fine tuned~\cite{RevModPhys.78.373}. Yet in FST the in-gap states at many of the magnetic impurities are observed at zero energy~\cite{yin2015observation,fan2021observation}. The other surprising property is the absence of a hybridization gap in the Shiba-state energies of two nearby magnetic impurities~\cite{yin2015observation}, which contrasts with the standard behavior seen in conventional systems~\cite{PhysRevLett.120.156803}. STM measurements have further shown that, while the energy of the in-gap state is zero when the impurity is at the center of the unit cell, it becomes finite when the impurity is pushed toward the edge of the unit cell~\cite{fan2021observation}. 
Proposals such as anomalous quantum vortices forming at magnetic impurities~\cite{PhysRevX.9.011033} or effective $\pi$-phase shifts at the impurity sites~\cite{newpi,bjornson2017superconducting} have been put forward, but a comprehensive description of properties of these states is still lacking. 

In this Letter we present an alternative mechanism for formation of zero energy states close to magnetic impurities in FST. The key property of FST leading to the type of in-gap states discussed in this Letter is the possible existence of multiple superconducting pairing instabilities energetically close to each other. Near an impurity, the Zeeman field generated by it strongly impacts this balance between different types of pairing. In particular, by solving the linearized gap equations, we find that the pairing state favored by the Zeeman field in this region is topologically distinct from the pairing state in regions farther than a multiple lattice spacing from the magnetic impurity. As a result, a pair of zero-energy states form in the boundary region between the two types of pairing. The resulting zero-energy states form bubbles surrounding the magnetic impurity, with a radius of the order of multiple lattice spacing. Since these states arise from the topological character of the superconducting state, their energy is generally pinned at zero. 

When two impurities approach each other, the regions around them where the topological superconductivity is dominant merge. As a result, the zero modes surround a larger area and do not become gapped. Furthermore, the type of pairing that is preferred close to the impurity is determined by the symmetries of the system. Therefore, the position of the impurity in the unit cell can affect the type of pairing selected and, consequently, the development of zero modes. Interestingly, the triplet superconducting pairing that forms close to the magnetic impurity is time-reversal symmetric, and the boundary states are Kramers pairs that are insensitive to the direction of the spin-polarization of the magnetic impurity. The later property is consistent with results from spin-polarized STM measurements~\cite{wang2021spin}. Another feature of the magnetic-impurity induced in-gap state in FST is that it becomes gapped once the magnetic impurity approaches a magnetic vortex~\cite{fan2021observation}. This is because the two zero modes enclosing the magnetic impurity hybridize in the presence of a Zeeman field. Therefore, due the vortex hosting a sizable magnetic field, it gaps out the zero modes.

\textit{Model Hamiltonian:} FST is a member of the family of iron-based superconductors \cite{Fernandes2022iron} with non-symmorphic $P4/nmm$ space group symmetry due to the buckling of the chalcogen atoms inside the $2$-Fe unit cell~\cite{fernandes2016low,cvetkovic2013space,lohani2020band}. At the $\Gamma$ point, the $P4/nmm$ space group is isomorphic to the $D_{4h}$ point group. The generators of the $D_{4h}$ group can be taken to be $\pi_x$, $\pi_z$ and $\pi_X$ reflection planes. Here, $x$ and $y$ connect nearest-neighbor Fe atoms, whereas $X$ and $Y$, rotated by $45^\circ$ with respect to $x$ and $y$, connect next-nearest-neighbor Fe atoms. The FeSe layers are stacked along $z$. The band structure close to $\Gamma$ point mainly contains $p_z$, $d_{xz}$ and $d_{yz}$ orbitals and the most general effective Hamiltonian including spin-orbit coupling (SOC) can be constructed using the method of invariants~\cite{inui2012group,bir1974symmetry}. The inversion-odd $p_z$ orbital is essential for realizing band inversion along the $\mathrm{\Gamma}-Z$ direction and thus the nontrivial topology of the band structure~\cite{xu2016topological,lohani2020band}. In the basis $\psi_\mathbf{k}=\left(|d_+\uparrow\rangle,|d_-\uparrow\rangle,|p_z\uparrow\rangle,|d_+\downarrow\rangle,|d_-\downarrow\rangle,|p_z\downarrow\rangle\right)$, where $d_\pm=\left(d_{yz}\pm id_{xz}\right)^T$, the Hamiltonian is given by~\cite{lohani2020band}
\begin{equation}
H\left(\mathbf{k}\right)=\sigma_0 \otimes H_0\left(\mathbf{k}\right)+H_\mathrm{SOC}\left(\mathbf{k}\right),
\end{equation} 
where for small in-plane momentum:
\begin{align}
H_0\left(\mathbf{k}\right)&=\left(\begin{array}{ccc}
M_1\left(\mathbf{k}\right) & \beta k^2_+ & \delta k_-\\
\beta k^2_-& M_1\left(\mathbf{k}\right) & -\delta k_+ 	\\
\delta k_+& -\delta k_-& M_2\left(\mathbf{k}\right)
\end{array}\right),
\end{align}
with $k_\pm=k_x\pm i k_y$ and $M_n\left(\mathbf{k}\right)=E_n+\left(k^2_x+k^2_y\right)/2m_{nx}+t_{nz}\left(1-\cos\left(k_z\right)\right)$. 

The non-zero elements of the SOC Hamiltonian $H_\mathrm{SOC}\left(\mathbf{k}\right)$ are
%\begin{align}
$H^\mathrm{22}_\mathrm{SOC}\left(\mathbf{k}\right)=H^\mathrm{44}_\mathrm{SOC}\left(\mathbf{k}\right)=-H^\mathrm{11}_\mathrm{SOC}\left(\mathbf{k}\right)=-H^\mathrm{55}_\mathrm{SOC}\left(\mathbf{k}\right)=\lambda_1$, and
$H^\mathrm{16}_\mathrm{SOC}\left(\mathbf{k}\right)=H^\mathrm{35}_\mathrm{SOC}\left(\mathbf{k}\right)=\sqrt{2}\lambda_3\sin\left(k_z\right)$, as well as the matrix elements related by hermiticity.
%and all other terms of $H_\mathrm{SOC}\left(\mathbf{k}\right)$ being zero or determined by hermiticitiy. 
The parameters $\beta$, $\delta$, $E_1$, $E_2$, $m_1$, $m_2$, $t_1$, $t_2$, $\lambda_1$ and $\lambda_3$
were previously determined using density functional theory~\cite{xu2016topological}.
\begin{figure}[t]
    \centering
    \includegraphics[width=0.99\columnwidth]{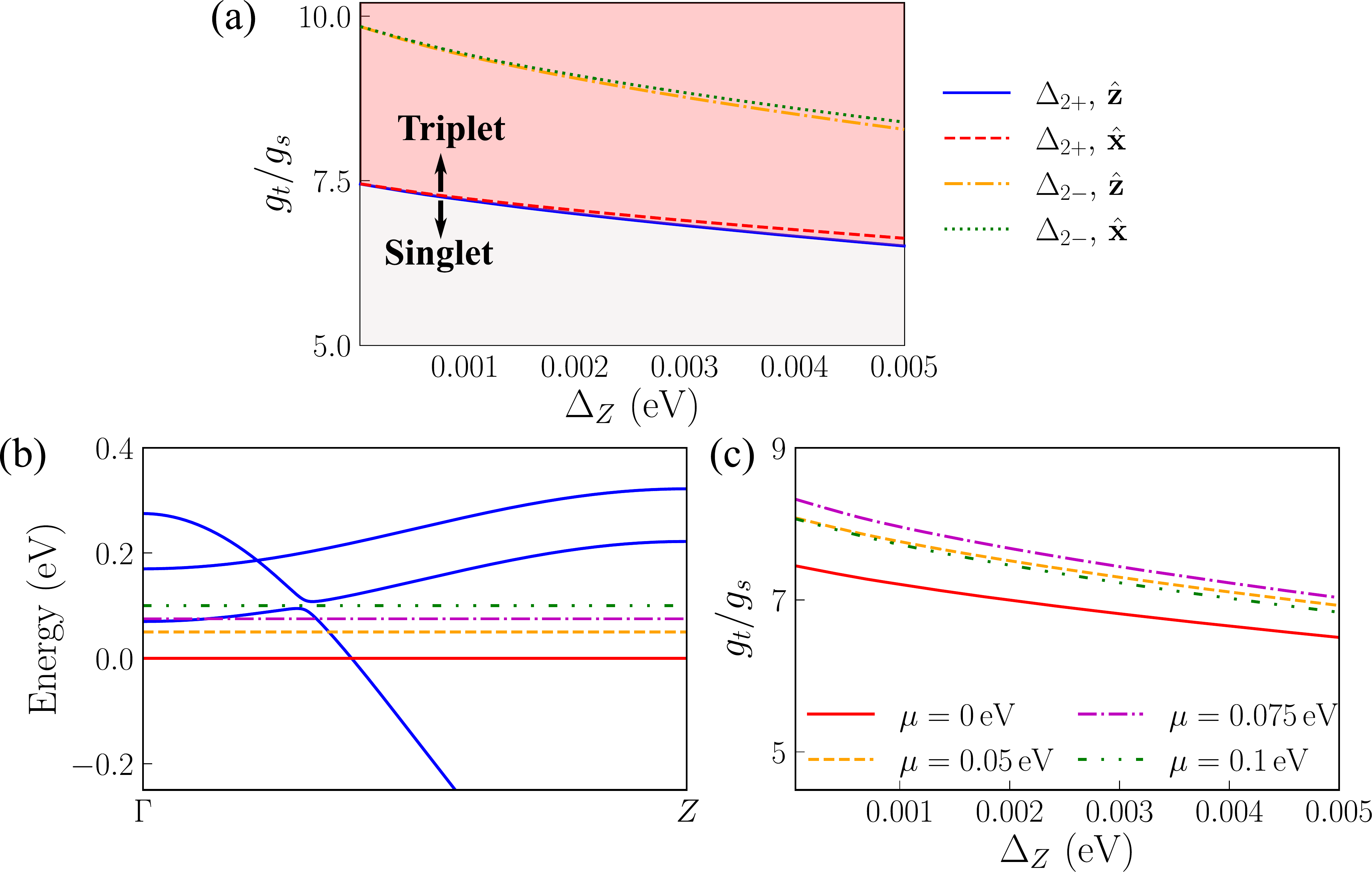}
    \caption{(a) Phase boundary between different inter-orbital triplet ($\Delta_{2\pm}$) and intra-orbital spin-singlet ($\Delta_1$) pairing states for $\mu=0\,\mathrm{eV}$ when the Zeeman field is along the $\hat{\mathbf{z}}$ or $\hat{\mathbf{x}}$ direction. Singlet (triplet) pairing emerges in the light gray (light red) region, following the phase boundary for the Zeeman field along the $\hat{\mathbf{z}}$ direction.   (b) Band structure in the $\Gamma-Z$ direction. The selected chemical potentials are those for which the phase boundaries between $\Delta_{2+}$ and $\Delta_1$ are shown in (c). The Zeeman field is in the $\hat{\mathbf{z}}$ direction for (c).}
    \label{fig1}
\end{figure}
\textit{Symmetries and pairing states: }
The types of pairing states in FST can be classified according to irreducible representations of the $D_{4h}$ symmetry group of the underlying lattice~\cite{RevModPhys.63.239}. $D_{4h}$ has five odd-parity and five even-parity irreducible representations. We only consider pairing with zero center-of-mass momentum. In the even sector, the standard pairing is intra-orbital spin-singlet that transforms as the $\mathrm{A_{1g}}$ representation and has the form $\Delta_1:\sigma_0\otimes \mathbf{1}_3$. We use the standard BdG basis where the wave function has the form $\Psi_\mathbf{k}=\left(\psi_\mathbf{k},i\sigma_yK\psi_\mathbf{k}\right)^T$. Due to the presence of SOC, there is a spin-triplet inter-orbital pairing involving the $d_{xz}$ and $d_{yz}$ orbitals that also transforms as $\mathrm{A_{1g}}$, of the form $\sigma_zd_{xz}d_{yz}$. This pairing state, favored by the Hund's coupling~\cite{vafek2017hund}, does not pair the $p_z$ orbitals. Therefore, when the chemical potential is close to the $p$-$d$ band inversion point, $\sigma_zd_{xz}d_{yz}$ pairing state is not energetically favorable. 

On the other hand, there are four types of odd-parity pairing states that are inter-orbital triplet and involve the $p_z$ orbital. They are $\Delta_{2\pm}:\sigma_xp_zd_{xz}\pm\sigma_yp_zd_{yz}$, which transforms as the $\mathrm{A_{1u}}$ ($+$) or $\mathrm{B_{1u}}$ ($-$) irreducible representation; and $\Delta_{3\pm}:\sigma_yp_zd_{xz}\mp\sigma_xp_zd_{yz}$, which transforms as $\mathrm{A_{2u}}$ ($+$) or $\mathrm{B_{3u}}$ ($-$). In the $d_\pm$ basis, these gap functions generate the Bogoliubov-de Gennes mean-field pairing terms:
\begin{align}
\label{Pair_1}
H_{\Delta_{\{2,3\},+}}&=\Delta_{\{2,3\},+}i^{\{0,1\}}\tau_x\left(\begin{array}{cccccc}
0 &0 &0 &0 &0 &1 \\
0 &0 &0 &0 &0 &0 \\
0 &0 &0 &0 &-1 &0 \\
0 &0 &0 &0 &0 &0 \\
0 &0 &\mp1 &0 &0 &0 \\
\pm1 &0 &0 &0 &0 &0 \\
\end{array}\right), \\
H_{\Delta_{\{2,3\},-}}&=\Delta_{\{2,3\},-}i^{\{0,1\}}\tau_x\left(\begin{array}{cccccc}
0 &0 &0 &0 &0 &0 \\
0 &0 &0 &0 &0 &1 \\
0 &0 &0 &-1 &0 &0 \\
0 &0 &\mp1 &0 &0 &0 \\
0 &0 &0 &0 &0 &0 \\
0 &\pm1 &0 &0 &0 &0 \\
\end{array}\right),
\label{Pair_2}
\end{align}
where the upper (lower) signs refer to $\Delta_2$ ($\Delta_3$) and the Pauli matrix $\tau_j$ acts in Nambu particle-hole space. The overall phase was chosen such that both $\Delta_{2,\pm}$ and $\Delta_{3,\pm}$ respect time-reversal symmetry. 

From numerical analysis, we identify that all these pairing states, except $\Delta_{2+}$, are nodal. The novel structure of the pairing gaps has important implications for the topological character of the superconducting state. Indeed, a time-reversal and odd-parity pairing is topological if it is gapped and encloses an odd number of time-reversal invariant momenta~\cite{fu2010odd}. The space group $P4/nmm$ has the non-symmorphic symmetry operation inversion followed by half translation, which at the $\Gamma$ point corresponds to the inversion symmetry operation of $D_{4h}$~\cite{cvetkovic2013space}. As a result, the triplet pairings of FST identified above, $\Delta_{2\pm}$ and $\Delta_{3\pm}$, have odd parity. Since only $\Delta_{2+}$ has a full gap, it is the only one among those four {that can be in a topological superconducting state, as long as it encloses an odd number of time-reversal invariant momenta}.   

\textit{Effect of the Zeeman field:} To determine the dominant pairing instability as a function of the Zeeman field we utilize the linearized gap equation \cite{Fischer_2013}. We start by defining the finite temperature superconducting susceptibilities:
\begin{align}
\chi_{lm}&=-\frac{T}{N}\sum_{\omega_n\mathbf{p}}\mathrm{Tr}\left[\frac{H_{\Delta_l}^\dagger}{\Delta_l}G_0\left(i\omega_n,\mathbf{p},\Delta_Z\right)\times\right. \nonumber \\
&\hspace{80pt}\left.\frac{H_{\Delta_m}}{\Delta_m}G_0\left(-i\omega_n,\mathbf{p},-\Delta_Z\right)\right],
\end{align}
where $H_{\Delta_l}$ is the pairing Hamiltonian, $N$ is the number of momentum points, and $G_0\left(i\omega_n,\mathbf{p},\Delta_z\right)=\sum_j\mathcal{P}_{j\mathbf{p}\Delta_Z}/\left(i\omega_n-\epsilon_{j\mathbf{p}\Delta_z}\right)$ is the normal state Green's function. $j$ runs through the bands, $\omega_n$ is the Matsubara frequencies and $\mathcal{P}_{j,\mathbf{p}\Delta_Z}$ is the projection operator onto band $j$ at momentum $\mathbf{p}$. The Zeeman-field energy splitting is $\Delta_Z$. Finally, the subscripts $l$ and $m$ label the five pairing states discussed above: the singlet $\Delta_1$ and the triplets $\Delta_{2\pm}$, $\Delta_{3\pm}$. The linearized gap equations, which determine the critical temperature $T_c$ for each pairing instability, take the form

\begin{align}
-\frac{1}{g_s}-\chi_{11}&=0, \\
-\frac{2}{3g_{t}}-\chi_{tt}&=0,
\end{align}
where $g_l$ are the superconducting coupling constants arising from the microscopic interactions, with $s=1$ and $t=\pm2, \, \pm3$. In deriving these expressions, we ignored mixing between different pairing channels, e.g. $\Delta_{2+}$ and $\Delta_{2-}$, that is allowed by the Zeeman field. We will justify this later. %\rmf{\emph{RMF: I commented out the next sentence.}}%Note that the $2/3$ prefactor for the triplet case is due to the fact that $\Delta_{2\pm}$ pair only limited orbitals with specific spin as can be seen in (\ref{Pair_1}) and (\ref{Pair_2}).  
\begin{figure}[t]
    \centering
    \includegraphics[width=0.99\columnwidth]{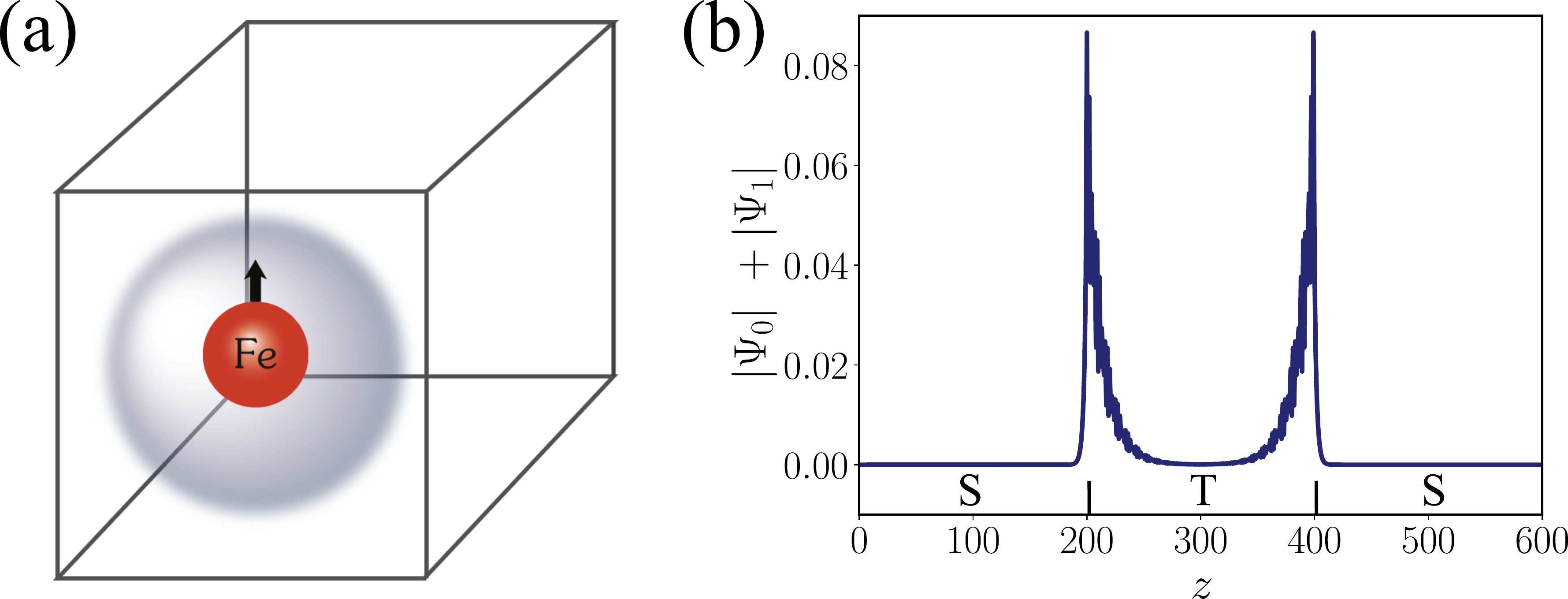}
    \caption{(a) Schematic representation of the interstitial Fe atom and its surrounding triplet pairing state, separated from the spin-singlet pairing realized away from the atom. (b) Corresponding total probability distribution, along the $\hat{\mathbf{z}}$ direction, for the zero-energy mode Kramers pair. The system size is 600 lattice sites, divided into three regions of singlet/triplet (topological)/singlet pairings with 200 sites each. The chemical potential is $\mu=0\,\mathrm{meV}$, and $k_x=k_y=0$.
    }
    \label{fig2}
\end{figure}
%The the linearized gap equation determines the critical temperature for each type of pairing and % for different values of coupling constant and
%the dominant pairing is associated with the highest critical temperature $T_c$.
%Once $T_c$ is determined linearized gap equation can also be used to identify the relative amplitude and phase of the $\Delta_2$ and $\Delta_3$ pairings. The later two pairings belong to separate irreducible representation in the absence of Zeeman field and their associated coupling constant can be different. The presence of the Zeeman field leads to their coupling. From the linearized gap equation, we could see that slight difference in the coupling associated with the two pairing could lead to strong suppression of $\Delta_3$ compared to $\Delta_2$. %While both pairings respect time-reversal symmetry the combined pairing generally does not.

To proceed, we need to discuss the coupling constants $g_l$. We assume, without specifying a mechanism, that there is attractive interaction in the singlet channel, since this is the state realized in the bulk. As for the triplet states, according to what we discussed above, only $\Delta_{2+}$ can correspond to a topological pairing state, which can display zero-energy modes. It is therefore crucial to identify the microscopic interactions that favors $g_{2\pm}$ over $g_{3\pm}$. We start from a generalized Hubbard-Kanamori interacting Hamiltonian \cite{kanamori,georgeshund,graser2009near} for the $p_z$, $d_{xz}$, and $d_{yz}$ orbitals, which includes anisotropic Hund's terms $J_1$, $J_2$ and $J_3$. We find a simple condition under which there is an effective attractive interaction only for the $\Delta_{2\pm}$ states (see Supplementary Information): $J_1>J_2$ and $J_3+J_1-J_2<V<J_3+J_2-J_1$, where $V$ is the inter-orbital Hubbard repulsion. We assume this condition is satisfied, set $g_{2\pm}=g_t$, and focus hereafter on the $\Delta_{2\pm}$ and $\Delta_1$ states only. 

%To accomplish that we first calculate superconducting susceptibility by discretizing momentum and carrying integration numerically. For that we transform into cylindrical coordinates and use the fact that the integrand is azimuthally symmetric. We take 400 points in radial directions in a range of $(0,0.3)$ and 800 points in $\hat{\mathbf{z}}$ for the full Brillouin zone. 

%We consider the effect of the Zeeman coupling $H_Z=\mathbf{\Delta}_Z\cdot\boldsymbol{\sigma}\otimes \mathbf{1}_3$ in the Hamiltonian $H\left(\mathbf{k}\right)$ on the preferred pairing. 
Fig.~\ref{fig1} (a) shows the phase boundaries for the different pairing states in the parameter space of Zeeman splitting and the coupling constants ratio $g_t/g_s$. The Zeeman field is taken in the $\hat{\mathbf{z}}$ or $\hat{\mathbf{x}}$ (in-plane) direction. The phase boundaries between $\Delta_{2+}$ and $\Delta_{2-}$ are quite separated, such that $\Delta_{2+}$ dominates for small values of the coupling constant ratio, regardless of the field direction. This not only shows that $\Delta_{2+}$ is the favored pairing state for larger Zeeman coupling, but also that the Zeeman-induced mixing with $\Delta_{2-}$ should be small, which justifies dropping this term in the gap equations. Fig.~\ref{fig1} (a) shows that phase boundary is almost insensitive to the direction of the Zeeman field. %Fig.~\ref{fig1} (b) presents $T_c$ as a function of coupling constants $g_s$ and $g_t$ for corresponding pairings with zero Zeeman field. The obtained $T_c$ is in the range of a few meV comparable to the bulk superconducting gap observed in the experiment \cite{yin2015observation,zhang2018observation}. 

%As can be seen in the figures, the phase boundary for $\Delta_{2+}$ and $\Delta_{3+}$ is quite close, but in $\hat{\mathbf{z}}$ direction for all considered values of Zeeman field $\Delta_{2+}$ has the smallest coupling constant (this is also confirmed by the difference between coupling constants of $\Delta_{3+}$ and $\Delta_{2+}$ shown in the inset). Therefore, it will be the dominant pairing for larger Zeeman coupling crossing the phase boundary. For Zeeman field in $\hat{\mathbf{x}}$ direction the difference between coupling constants of $\Delta_{3+}$ and $\Delta_{2+}$ changes sign with the increase of Zeeman field. Since $\Delta_{3+}$ pairing is not topological the presence of Majorana modes around iron impurity discussed below should be modified with the increase of in plane Zeeman field. 

To show that $\Delta_{2+}$ pairing is robust with respect to changes in the chemical potential, Fig.~\ref{fig1} (b,c) shows the small changes experienced by the $\Delta_{2+}$-$\Delta_1$ phase boundary for four different values of $\mu$. Therefore, upon increasing the Zeeman coupling, the pairing state will transition from intra-orbital singlet $\Delta_1$ to inter-orbital triplet $\Delta_{2+}$. 

While the preferred pairing is robust with respect to the chemical potential modification, its topological nature depends on the chemical potential. The relevant time-reversal invariant momenta are at the $\Gamma$ and $Z$ points. For $\mu>-0.577\,\mathrm{eV}$, the Fermi surface at the $Z$ point is filled whereas for $\mu<0.07\,\mathrm{eV}$, all three bands at $\Gamma$ are empty. Therefore, in this range of $\mu$, the $\Delta_{2+}$ pairing state is topological. Consequently, in Fig.~\ref{fig1} (c,d), only two chemical potential values $\mu=0\,\mathrm{eV}$ and $\mu=0.05\,\mathrm{eV}$ correspond to a topological superconductor. The cases $\mu=0.075\,\mathrm{eV}$ and $\mu=0.1\,\mathrm{eV}$ are generally non-topological. For $\mu=0.075\,\mathrm{eV}$, the lowest band at the $\Gamma$ point is quite close to the Fermi energy. As a result, a topological phase transition can be induced by increasing the pairing amplitude or further tuning the chemical potential. Since this requires considerable tuning, difficult to achieve experimentally, we focus on values of the chemical potential where $\Delta_{2+}$ is strictly topological.

When the chemical potential is in the band-inversion gap, the normal-state itself is topological~\cite{xu2016topological} (see Fig.~\ref{fig1} (c)). This is manifested by Majorana bound states at the end of the vortex cores where they cross the sample boundary~\cite{fukanemajorana,wang2018evidence}. In this case, the bulk pairing state is intra-orbital spin-singlet $\Delta_{1}$ which is not topological in the bulk, but it also induces topological superconductivity on the surface~\cite{fukanemajorana}. Since our interest is in the vicinity of bulk impurities, we will not discuss surface effects.

\begin{figure}[t]
    \centering
    \includegraphics[width=0.99\columnwidth]{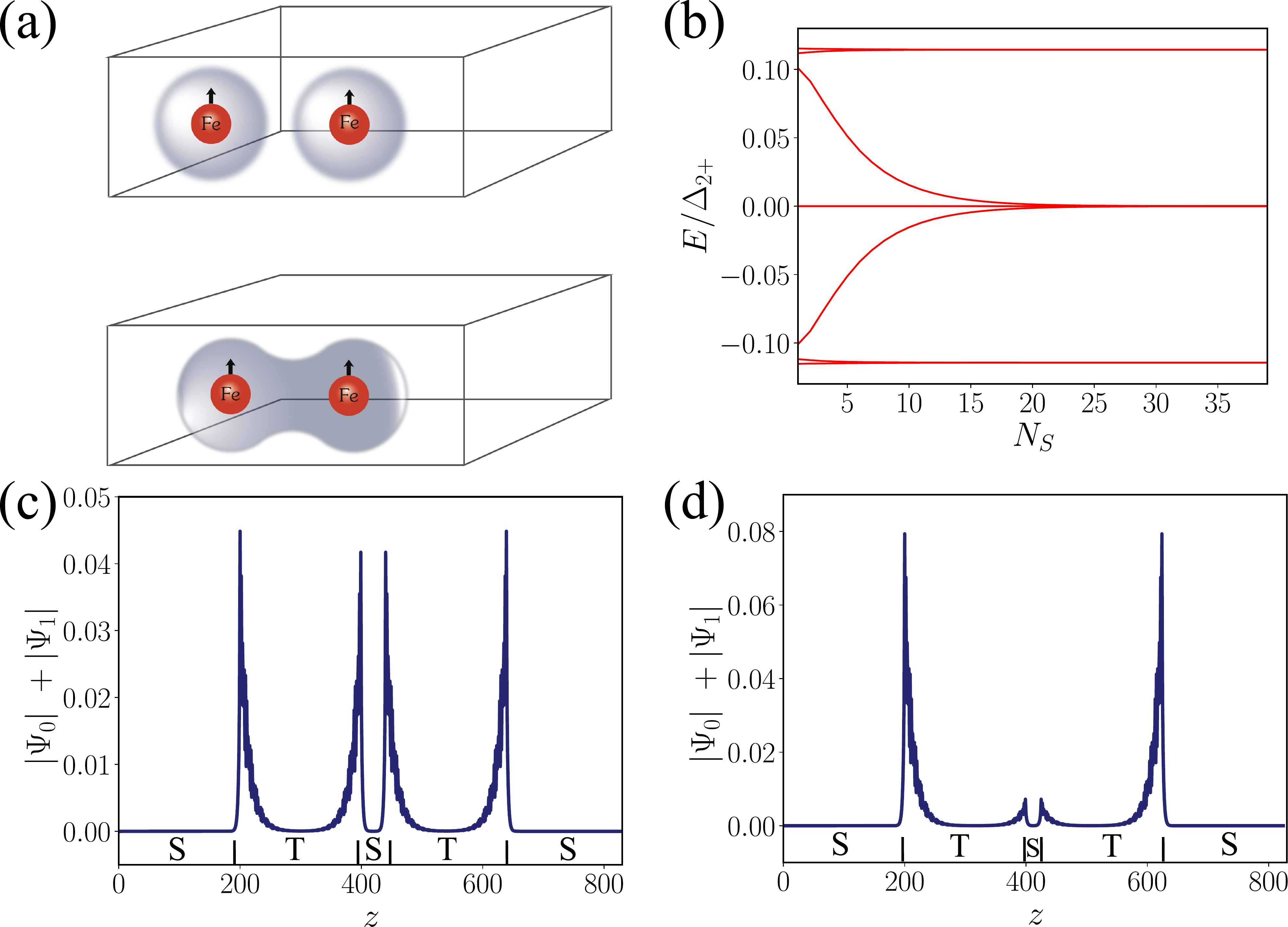}
    \caption{(a) Schematic representation of bringing two interstitial Fe atoms closer. (b) Dependence of the in-gap energy states on the effective two impurity separation in $\hat{\mathbf{z}}$ direction. The two impurity state is simulated by a five region sandwich with superconducting pairings singlet/triplet/singlet/triplet/singlet; the distance between the boundaries of the two triplet regions is $N_S$ sites. (c,d) Corresponding total probability when two Fe atoms are farther apart and closer to each other, respectively. For (c,d) the system is the same as in (b) with $N_S=25$ for (c) and $N_S=40$ for (d). Chemical potential is $\mu=0\,\mathrm{meV}$ and $k_x=k_y=0$.
    }
    \label{fig3}
\end{figure}

\textit{Majorana modes:}
Having identified the dominant pairing state in the presence of Zeeman coupling, we now examine the possible boundary mode localized between regions with $\Delta_{2+}$ and spin-singlet $\Delta_1$ pairings. Since for a considerable range of chemical potential values the $\Delta_{2+}$ pairing is topological, the presence of zero-energy edge modes is expected. The $\Delta_{2+}$ pairing respects time-reversal symmetry, therefore the zero modes are Kramers pairs with opposite spin. %It should be noted that the magnetic field breaks time-reversal symmetry, thus splitting these two pairs. 

Figs.~\ref{fig2} and \ref{fig3} depict the structure of zero modes for single and double magnetic impurities. Figs.~\ref{fig2} (a) and (b) show the case of a single Fe impurity. The effective magnetic field generated by the magnetic impurity modifies the superconducting order parameter in its surroundings, such that $\Delta_{2+}$ pairing is realized. The boundary with the bulk singlet pairing features zero energy modes. Therefore, scanning tunneling spectroscopy close to the interstitial iron atom should detect a robust zero bias peak~\cite{yin2015observation,fan2021observation,wang2021spin}. % similar to that seen in the vortex cores \cite{wang2018evidence}.
The observed zero mode is different from usual Shiba state observed next to a magnetic impurity \cite{balatsky2006impurity}, whose energy depends on the amplitude of the exchange interaction and requires special tuning to be fixed at zero energy. In contrast, in our case, as long as the magnetic field is sufficiently large to favor topological $\Delta_{2+}$ pairing, the zero-energy mode is robust. 

Fig.~\ref{fig3} shows the evolution of the zero modes when the two impurities are brought close to each other. As shown in Fig.~\ref{fig3} (b), while two of the zero modes are gapped out, the ones located on the outer edge of the two-impurity region remain robust. This is consistent with the experimental observation in Ref.~\cite{yin2015observation} and is different from the expected hybridization between Shiba states when the two impurities are brought close to each other. {This shows that the corresponding zero modes form bubbles around the magnetic impurities, which combine into a single bubble enclosing both atoms when the impurities are close to each other}. %Majorana vortex modes when two vortices are brought close to each other \cite{chiu2020scalable}.    

{Finally, we comment on the size of the bubble formed around the interstitial magnetic impurity. Analyzing the Friedel oscillations from neutron scattering data, Ref. \cite{PhysRevLett.108.107002} used a five-orbital Hubbard model to estimate that the nearest-neighbor spin exchange between the interstitial and the surrounding Fe atoms is about 70 meV. This is an order of magnitude larger than the Zeeman field required to observe the triplet pairing state and the topological superconducting region around the magnetic impurity (see Fig.~\ref{fig1}). Therefore, despite the exchange interaction being short-ranged, we expect the radius of the bubble to be in the range of a multiple lattice spacing, consistent with experimental observations~\cite{yin2015observation}}.

\textit{conclusions:} We have shown that the peculiar features of the zero bias peak experimentally observed at interstitial iron atoms in FST can be reconciled if we consider the modification of the superconducting order parameter close to the impurity. We have shown that the Zeeman field prefers inter-orbital triplet pairing which, for a certain range of chemical potential values, is topological. As a result, zero-energy modes naturally occur at the boundary between inter-orbital triplet and intra-orbital singlet pairings. These modes are robust and do not get modified by changes in the exchange interaction, contrary to the conventional Shiba states.  It has been experimentally observed that the zero modes are not spin-polarized \cite{wang2021spin}, a feature that is also consistent with our theoretical model. The obtained triplet pairing state near the impurity respects time-reversal symmetry, such that the zero modes are always doubly-degenerate and have opposite spin due to the Kramers theorem. This mechanism is also capable of explaining the robustness of the zero modes when two impurities are brought close to each other, since in this case only two out of four modes get hybridized.   

\begin{acknowledgements}
{We thank Armin Rahmani, Andrey V. Chubukov, Jay D. Sau and Ruixing Zhang for fruitful discussions. AK and PG are supported by NSF-DMR2037996. PG also acknowledges support from NSF-DMR1824265. RMF was supported by the U. S. Department of Energy, Office of Science, Basic Energy Sciences, Materials Sciences and Engineering Division, under Award No. DE-SC0020045. Part of this work was performed at the Aspen Center for Physics, which is supported by National Science Foundation grant PHY-1607611.} 
\end{acknowledgements}

\bibliography{shiba_refs}
\end{document}

\renewcommand{\thefigure}{S\arabic{figure}}
\renewcommand{\figurename}{Figure}
\setcounter{figure}{0}
\setcounter{equation}{0}
\renewcommand{\theequation}{S\arabic{equation}}

\newcommand\rmf[1]{\textcolor{magenta}{#1}}

\clearpage

\onecolumngrid

\begin{center}
	\textbf{\large Supplementary material for: \strut\\ ``\mytitle'' }\\[5pt]
	% 	\begin{quote}
	% 		{\small 
	% 			In this supplementary material we  }\\[20pt]
	% 	\end{quote}
\end{center}
\author{Areg Ghazaryan}
\affiliation{Institute of Science and Technology Austria, Am Campus 1, 3400 Klosterneuberg, Austria}

\author{Ammar Kirmani}
\affiliation{Physics Department, City College of the City University of New York, NY 10031, U.S.A.}

\author{Rafael M. Fernandes}
\affiliation{School of Physics and Astronomy, University of Minnesota, Minneapolis, Minnesota 55455, US}

\author{Pouyan Ghaemi}
\affiliation{Physics Department, City College of the City University of New York, NY 10031, U.S.A.}
\affiliation{Physics Program, Graduate Center of City University of New York, NY 10031, U.S.A.}
\maketitle
\section*{Pairing from interacting Hamiltonian}
Here we establish the relationship between the superconducting coupling constants $g_{2\pm}$ and $g_{3\pm}$ of the odd-parity pairing states discussed in the main text and the onsite interaction terms of a generalized Hubbard-Kanamori Hamiltonian \cite{kanamori,georgeshund,graser2009near} for the three orbitals $d_{xz}$, $d_{yz}$ and $p_z$:
\begin{align}
&H_{int}=U\sum_{i\mu}n_{i\mu\uparrow}n_{i\mu\downarrow}+\frac{V}{2}\sum_{i\mu\mu^\prime\neq\mu}n_{i\mu}n_{i\mu^\prime}-\frac{J_1}{2}\left(\sum_{i,\mu\neq\mu^\prime={d_{xz},p_z}}S_{x,i\mu}S_{x,i\mu^\prime}+\sum_{i,\mu\neq\mu^\prime={d_{yz},p_z}}S_{y,i\mu}S_{y,i\mu^\prime}\right)-\nonumber \\
&\frac{J_2}{2}\left(\sum_{i,\mu\neq\mu^\prime={d_{xz},p_z}}S_{y,i\mu}S_{y,i\mu^\prime}+\sum_{i,\mu\neq\mu^\prime={d_{yz},p_z}}S_{x,i\mu}S_{x,i\mu^\prime}\right)-\frac{J_3}{2}\left(\sum_{i,\mu\neq\mu^\prime={d_{xz},p_z}}S_{z,i\mu}S_{z,i\mu^\prime}+\sum_{i,\mu\neq\mu^\prime={d_{yz},p_z}}S_{z,i\mu}S_{z,i\mu^\prime}\right)-\nonumber \\
&\frac{J_{d}}{2}\left(\sum_{i,\mu\neq\mu^\prime={d_{xz},d_{yz}}}S_{x,i\mu}S_{x,i\mu^\prime}+\sum_{i,\mu\neq\mu^\prime={d_{xz},d_{yz}}}S_{y,i\mu}S_{y,i\mu^\prime}\right)-\frac{J_{dz}}{2}\sum_{i,\mu\neq\mu^\prime={d_{xz},d_{yz}}}S_{z,i\mu}S_{z,i\mu^\prime}+\frac{J^\prime}{2}\sum_{i\mu\neq\mu^\prime s}\mu^\dagger_{is}\mu^\dagger_{i\bar{s}}\mu^{\prime}_{i\bar{s}}\mu^{\prime}_{is},
\label{HubbardHundHam}
\end{align}
where $U$ and $V$ describe the intra- and inter-orbital Hubbard interactions; $J_1$, $J_2$, $J_3$ describe the Hund's interactions between $d_{xz}$ and $p_z$ or $d_{yz}$ and $p_z$ orbitals; $J_d$ and $J_{dz}$ denote the Hund's interactions between the $d_{xz}$ and $d_{yz}$ orbitals; and $J^\prime$ describes the pair-hopping interaction. Here $n_{i\mu}=n_{i\mu\uparrow}+n_{i\mu\downarrow}$, $\bar{s}$ denotes the opposite spin to $s$, and the density and spin operators are defined as usual
\begin{align}
n_{i\mu s}&=\mu^\dagger_{is}\mu_{is}, \\
\mathbf{S}_{i\mu}&=\frac{1}{2}\sum_{s,s^\prime}\mu^\dagger_{i,s}\boldsymbol{\sigma}_{ss^\prime}\mu_{i,s^\prime}.
\end{align}
Note that, in Eq. (\ref{HubbardHundHam}), we consider the most general Hund's interactions allowed by $C_4$ symmetry and spin-orbit coupling. As a result, there are different Hund's couplings depending on the spin polarization and on the orbitals involved. This is necessary to distinguish between the four odd-parity superconducting states. Their gap functions can be written explicitly as 

\begin{align}
A_{1u}&:\Delta_{2+}=g_{2+}\sum_{s,s^\prime}\langle p_{zs}\left(i\sigma_x\sigma_y\right)_{ss^\prime}d_{xz,s^\prime}+p_{z,s}\left(i\sigma_y\sigma_y\right)_{ss^\prime}d_{yz,s^\prime}\rangle=g_{2+}\sum_{s,s^\prime}\langle -p_{zs}\sigma_{z,ss^\prime}d_{xz,s^\prime}+ip_{z,s}\sigma_{0,ss^\prime}d_{yz,s^\prime}\rangle, \\
B_{1u}&:\Delta_{2-}=g_{2-}\sum_{s,s^\prime}\langle p_{zs}\left(i\sigma_x\sigma_y\right)_{ss^\prime}d_{xz,s^\prime}-p_{z,s}\left(i\sigma_y\sigma_y\right)_{ss^\prime}d_{yz,s^\prime}\rangle=g_{2-}\sum_{s,s^\prime}\langle- p_{zs}\sigma_{z,ss^\prime}d_{xz,s^\prime}-ip_{z,s}\sigma_{0,ss^\prime}d_{yz,s^\prime}\rangle, \\
A_{2u}&:\Delta_{3+}=g_{3+}\sum_{s,s^\prime}\langle p_{zs}\left(i\sigma_y\sigma_y\right)_{ss^\prime}d_{xz,s^\prime}-p_{z,s}\left(i\sigma_x\sigma_y\right)_{ss^\prime}d_{yz,s^\prime}\rangle=g_{3+}\sum_{s,s^\prime}\langle i p_{zs}\sigma_{0,ss^\prime}d_{xz,s^\prime}+p_{z,s}\sigma_{z,ss^\prime}d_{yz,s^\prime}\rangle, \\
B_{2u}&:\Delta_{3-}=g_{3-}\sum_{s,s^\prime}\langle p_{zs}\left(i\sigma_y\sigma_y\right)_{ss^\prime}d_{xz,s^\prime}+p_{z,s}\left(i\sigma_x\sigma_y\right)_{ss^\prime}d_{yz,s^\prime}\rangle=g_{3-}\sum_{s,s^\prime}\langle i p_{zs}\sigma_{0,ss^\prime}d_{xz,s^\prime}-p_{z,s}\sigma_{z,ss^\prime}d_{yz,s^\prime}\rangle,
\end{align}
which shows that all gap functions involve pairs with same spins. Based on this, we split the interacting Hamiltonian in two parts
\begin{equation}
H_{int}=H_{0,int}+H^\prime_{int},
\end{equation}
where only $H_{0,int}$ incorporates same-spin pairing terms
\begin{align}
H_{0,int}&=\left(V-J_3\right)\sum_{i}\left(d^\dagger_{xz,i\uparrow}d_{xz,i\uparrow}p^\dagger_{z,i\uparrow}p_{z,i\uparrow}+d^\dagger_{xz,i\downarrow}d_{xz,i\downarrow}p^\dagger_{z,i\downarrow}p_{z,i\downarrow}\right)\nonumber \\
&\hspace{60pt}-\left(J_1-J_2\right)\sum_{i}\left(d^\dagger_{xz,i\uparrow}d_{xz,i\downarrow}p^\dagger_{z,i\uparrow}p_{z,i\downarrow}+d^\dagger_{xz,i\downarrow}d_{xz,i\uparrow}p^\dagger_{z,i\downarrow}p_{z,i\uparrow}\right) \nonumber \\
&\hspace{120pt}+\left(V-J_3\right)\sum_{i}\left(d^\dagger_{yz,i\uparrow}d_{yz,i\uparrow}p^\dagger_{z,i\uparrow} p_{z,i\uparrow}+d^\dagger_{yz,i\downarrow}d_{yz,i\downarrow}p^\dagger_{z,i\downarrow}p_{z,i\downarrow}\right)\nonumber \\
&\hspace{180pt}+\left(J_1-J_2\right)\sum_{i}\left(d^\dagger_{yz,i\uparrow}d_{yz,i\downarrow}p^\dagger_{z,i\uparrow}p_{z,i\downarrow}+d^\dagger_{yz,i\downarrow}d_{yz,i\uparrow}p^\dagger_{z,i\downarrow}p_{z,i\uparrow}\right),
\end{align}
In contrast, $H^\prime_{int}$ contains only opposite-spin pairing terms, and is thus not relevant for the current treatment. From the interacting Hamiltonian, it is now straightforward to read off the coupling constants for each superconducting channel:
\begin{align}
g_{2+}+g_{2-}+g_{3+}+g_{3-}&=J_3-V, \\
g_{2+}+g_{2-}-g_{3+}-g_{3-}&=J_2-J_1.
\end{align}
We therefore obtain
\begin{align}
g_{2+}+g_{2-}&=\frac{-V+J_3-J_1+J_2}{2}, \\
g_{3+}+g_{3-}&=\frac{-V+J_3+J_1-J_2}{2}.
\end{align}
Upon requiring the $A_{1u}$ and $B_{1u}$ channels to be attractive ($g_2>0$) and the $A_{2u}$ and $B_{2u}$ channels to be repulsive ($g_3<0$), we find the following condition
\begin{equation}
J_3+J_1-J_2<V<J_3+J_2-J_1,
\end{equation}
which can be satisfied if $J_1<J_2$ (assuming, as usual, $V>0$, $J_i>0$). If we further assume that $J_1$, $J_2$ and $J_3$ are of the same order $\approx J$, we see that, for this condition to hold, $V$ should generally be of the same order as $J$. This situation resembles the condition $J>V$ obtained in Ref.~\cite{vafek2017hund} to observe $s-$wave spin-triplet pairing. In this regard, it should be noted that $V$ and $J$ here are not the ``atomic" values, but the effective, low-energy values renormalized by high-energy degrees of freedom.

%, although in our case both $J>V$ and $J<V$ regimes are possible. We should note that generally $V>J$ based on first-principle calculations \cite{miyake2010comparison}.

We note that, in order to distinguish between $g_{2+}$ and $g_{2-}$, additional subleading terms would need to be included in the interacting Hamiltonian. These terms are enabled by the spin-orbit coupling, and thus correspond to terms of the form $\mathbf{L}_{i}\cdot\mathbf{S}_i$, where $\mathbf{L}_{i}=\sum_{\alpha\beta s s^\prime}d^\dagger_{i,\alpha,s}\tau_{y,\alpha\beta}\boldsymbol{\sigma}_{s s^\prime}d_{i,\beta,s^\prime}$ and $\tau$ acts on the $(d_{xz},d_{yz})$ basis. Since these terms are supposed to be subleading due to the moderate value of spin-orbit interaction in the system, we disregard them in our analysis. %While this will preclude us from distinguishing all inversion odd pairings, we still will be able to consider the role of Hubbard-Hund interaction amplitudes for competing and energetically close superconducting phases.
\bibliography{shiba_refs}